\documentclass[%
 reprint,
 amsmath,amssymb,
 aps,
]{revtex4-2}

\usepackage{graphicx}
\usepackage{dcolumn}
\usepackage{bm}
\usepackage{subcaption}
\usepackage{xcolor}
\usepackage[normalem]{ulem} 

\begin{document}

\preprint{APS/123-QED}

\title{A Sensitive Test of Non-Gaussianity in Gravitational-wave Detector Data}

\author{Ronaldas Macas}
 \email{ronaldas.macas@port.ac.uk}
\author{Andrew Lundgren}
 \email{andrew.lundgren@port.ac.uk}
\affiliation{
 Institute of Cosmology and Gravitation\\
 University of Portsmouth, Portsmouth, UK
}

\date{\today}

\begin{abstract}
Methods for parameter estimation of gravitational-wave data assume that detector noise is stationary and Gaussian.
Real data deviates from these assumptions, which causes bias in the inferred parameters and incorrect estimates of the errors.
We develop a sensitive test of non-Gaussianity for real gravitational-wave data which measures meaningful parameters that can be used to characterize these effects.
As a test case, we investigate the quality of data cleaning performed by the LIGO-Virgo-KAGRA collaboration around GW200129, a binary black hole signal which overlapped with the noise produced by the radio frequency modulation.
We demonstrate that a significant portion of the non-Gaussian noise is removed below 50\,Hz, yet some of the noise still remains after the cleaning; at frequencies above $\sim 85$\,Hz, there is no excess noise removed.
We also show that this method can quantify the amount of non-Gaussian noise in continuous data, which is useful for general detector noise investigations.
To do that, we estimate the difference in non-Gaussian noise in the presence and absence of light scattering noise.
\end{abstract}

\maketitle

\section{Intro}

Gravitational-wave (GW) data are often non-Gaussian meaning it contains outliers.
This excess noise is broadly divided into two categories: short and well-defined noise (also called `glitches'), and longer-duration broadband excess noise.
Both types of excess noise can overlap with real GW signals in time and frequency, therefore affecting GW searches, and source parameter estimation, including the sky localisation which is used for electromagnetic follow-up efforts~\citep{glitch_pe_1, glitch_pe_2, glitch_pe_3, gw170817_bw, skyloc}.

In some cases, the excess noise is recorded by the detector's witness channels.
For example, throughout the second LIGO-Virgo observing run multiple noise couplings were recorded: mains power lines, detector calibration lines, as well as the laser beam jitter noise present in LIGO Hanford~\citep{o1_o2_cat, o2_detchar}. 

In principle, recording the noise with witness channels allows us to estimate the noise coupling to the gravitational-wave strain channel and thus remove the noise~\cite{Nguyen_2021}.
With this in mind, the linear noise subtraction method \texttt{gwsubtract} was developed~\cite{davis_linsub}.
The method assumes that noise is Gaussian and stationary, and that the noise couples linearly.
About $30$\% increase in LIGO Hanford sensitivity was achieved during the second LIGO-Virgo observing run with \texttt{gwsubtract}~\cite{davis_linsub}. 
There was no noticeable improvement for LIGO Livingston because the `jitter' noise was not present at the detector.

During the third LIGO-Virgo-KAGRA observing run, binary black hole signal GW200129 coincided with the noise caused by the 45\,MHz electro-optic modulator driver system at LIGO Livingston~\cite{o3b_cat}.
This noise was recorded by multiple radio frequency (RF) channels, one of which was used to clean the data around GW200129 using the previously mentioned linear subtraction method~\cite{Davis_2022}.
After the re-analysis of LIGO-Virgo-KAGRA results, Hannam et al.\,(2022) found GW200129 to be a highly precessing event; in fact, the measured orbital precession is 10 orders of magnitude higher than any previous weak-field observation~\cite{Hannam_2022}.

Payne et al.\,(2022) questioned whether this event is indeed highly precessing or whether the observed precession is just an artefact of the left-over noise after the data cleaning~\cite{Payne_2022}. 
While there were various checks performed by the LIGO-Virgo-KAGRA collaboration to make sure that the linear noise subtraction mitigated the noise~\cite{o3b_cat}, there was no statistical estimate of how much RF noise was removed.

In this paper, we present a sensitive statistical test for measuring non-Gaussian noise in GW data.
In \S\ref{sec:methods}, we derive the normalised Q-transform which allows us to make assumptions about Gaussian data.
We also describe how Bayesian statistical modelling can be used to estimate the amount of non-Gaussian noise in GW data.
In \S\ref{sec:results}, we apply our method to the data around GW200129 and evaluate the effectiveness of the linear noise subtraction.
In addition, we also show that our method can be used as an informative real-time noise measurement.
Finally, the discussion and conclusions are given in \S\ref{sec:disc} and \S\ref{sec:conc}, respectively.

\section{Methods}
\label{sec:methods}
\subsection{Normalised Q-transform}
\label{sub:q_transform}
GW data containing outliers can be represented by the sum of Gaussian and non-Gaussian time series.
In our model, we assume that the Gaussian data follows a Gaussian distribution with a standard deviation of $1$, while the non-Gaussian noise is made out of short bursts of transient noise whose amplitudes are drawn from a distribution.

To identify Gaussian and non-Gaussian parts of the time series data, we start with the Q-transform defined as
\begin{equation}
    \label{eq:q_transform}
    X(\tau, f, Q) = \int_{-\infty}^{+\infty} x(t) w(t -\tau, f, Q) e^{-i2\pi f t} dt,
\end{equation}
where $w(t-\tau, f, Q)$ is a window centered on time $\tau$ with the width of the window proportional to the quality factor $Q$
\begin{equation}
    \label{eq:q}
    Q = \frac{f_0}{\Delta f},
\end{equation}
where $f_0$ is central frequency and $\Delta f$ is the frequency bandwidth~\cite{q_transform}.

Using Eq.\,(\ref{eq:q_transform}), we transform the time-series data into time-frequency tiles for certain $Q$ and $f_0$ values. Contrary to Ref.~\cite{Vazsonyi_2023} which uses a variable $Q$, we chose to use a specific window width for two reasons.

First, we focus only on frequencies where there is non-Gaussianity which means that our statistical measurement of noise is more precise.
Furthermore, by specifying $Q$ and $f_0$ we normalise the Q-tiles in such a way that the average power for Gaussian data in each tile $k$ is 1, i.e.
\begin{equation}
    \label{eq:power}
    \frac{1}{N} \sum_{k=1}^{k=N} \bigg[\Re^2\big(A_k(t)\big) + \Im^2\big(A_k(t)\big)\bigg] = 1,
\end{equation}
where $A(t)$ is the amplitude of time series data.

\begin{figure}
    \centering
    \includegraphics[width=0.4\paperwidth]{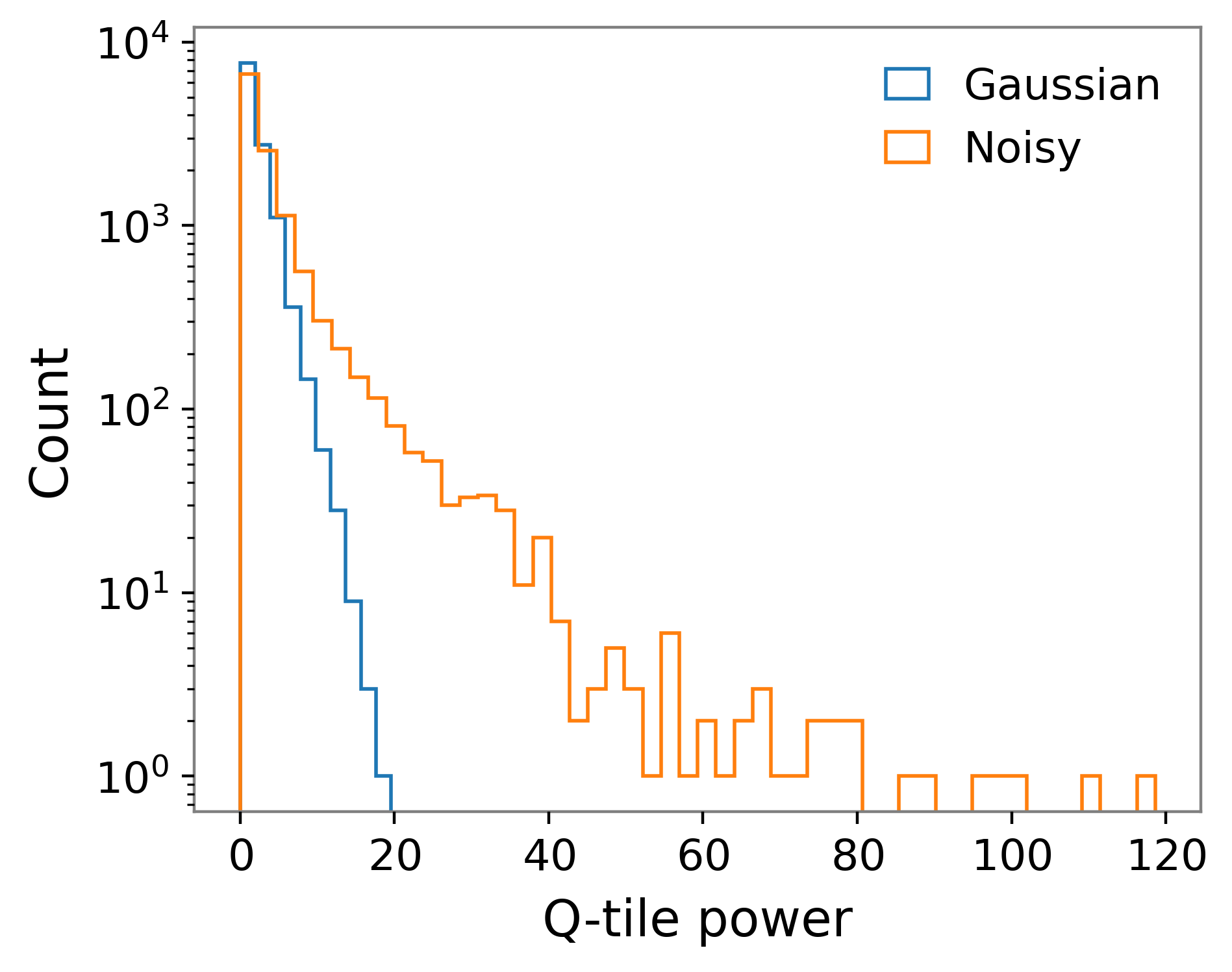}
    \caption{Q-tile power for $Q=8$ and $f_0=40$\,Hz for 366\,s simulated Gaussian data (blue) and 366\,s data with excess noise (orange).
    The distribution of powers for Gaussian data closely follows the expected exponential distribution with a rate $\lambda=1/2$.
    Noisy data, however, contains excess noise and therefore does not follow the exponential distribution associated with the Gaussian data.}\label{fig:power_hists}
\end{figure}

Normalising the Q-transform enables us to make an assumption about the distribution of Q-tile powers for Gaussian data.
As shown in Eq.\,(\ref{eq:power}), our Q-tiles have two degrees of freedom: real and imaginary parts of the time-frequency data.
As a consequence, we expect that Gaussian data should follow a $\chi^2$(2) distribution~\cite{chi_sq} which is simply an exponential distribution with a rate $\lambda=1/2$ (Figure \ref{fig:power_hists}).

This lets us define a simple statistical test to estimate the Gaussianity of the data: for the number of tiles $N$, the total Q-tiles power for Gaussian data is N and the standard deviation is $\sqrt{4N}$ (assuming $N \gg1$). 
\subsection{Distribution fitting with Bayesian statistical modelling}
\label{sub:pymc}
For a more informative test of non-Gaussianity, we model the non-Gaussian part of the data.
GW data with excess power has two contributions: Gaussian data modelled by $\chi^2(2)$ and a non-Gaussian part represented by an unknown distribution.
Data with excess power has much higher energies and a longer tail in contrast with the Gaussian-only data (Figure \ref{fig:power_hists}).

To estimate the amount of non-Gaussianity, we need to fit a mixture of two distributions.
We achieve this by using a marginalized mixture model likelihood~\cite{DFM}.

The likelihood of a data point $z_k$ belonging to either Gaussian data or non-Gaussian data distribution can be written as
\begin{multline}
    \label{eq:like}
    p(z_k|q_k, \alpha) = \Pi_{k=1}^{K} \big[ q_k p_1(z_k|\alpha_1) \\ + (1-q_k)p_2(z_k|\alpha_2)\big],
\end{multline}
where $k$ is the tile number, $q_k$ is a binary flag and $p_1(z_k|\alpha_1)$, $p_2(z_k|\alpha_2)$ represents the probability that a point corresponds to the Gaussian noise distribution or a non-Gaussian distribution, respectively.

Prior on $q_k$ is defined as
\begin{equation}
    \label{eq:prior}
    p({q_k}) =
    \begin{cases}
      F & \text{if $q_k$ = 0}\\
      1 - F & \text{if $q_k$ = 1}
    \end{cases}       
\end{equation}
where fraction $F \in [0, 1]$.

Eq.\,\eqref{eq:prior} allows us to marginalize the likelihood from Eq.\eqref{eq:like} giving
\begin{multline}
    \label{eq:marg}
    p(z_k, \alpha) = \Pi_{k=1}^K \big[(1-F)\,p_1(z_k|\alpha_1, q_k=1) \\ + F\,p_2(z_k|\alpha_2, q_k=0)\big].
\end{multline}
To estimate the marginalized likelihood, we use a Bayesian statistical modelling package PyMC with Hamiltonian Monte Carlo ``No-U-Turn" sampler (NUTS)~\citep{pymc, nuts}.
Rather than relying on classical Markov Chain Monte Carlo that does not consider the geometry of parameter space, PyMC with NUTS allows us to recover distribution parameters faster and with fewer samples. 
It achieves this by specifying a probabilistic model in PyMC, compiling it into a computational graph, providing initial values for parameters, configuring the NUTS sampler, and then running the sampling process.
NUTS explores the posterior distribution by proposing and accepting/rejecting parameter values, allowing for efficient and scalable Bayesian inference.

After recovering parameters for each distribution, we estimate the total power in each distribution, i.e.\,the area under the curve. 
The ratio of non-Gaussian power to the total power (which we refer to as \textit{fractional power}) lets us statistically measure non-Gaussianity in GW detector data. 

\section{Demonstration on real data}
\label{sec:results}
\subsection{Radio frequency noise around GW200129}%
\label{sub:gw200129}

GW200129 coincided with the $45$\,MHz radio frequency noise caused by the electro-optic modulator driver system at LIGO Livingston.
Luckily, this noise was also recorded by multiple witness channels, thus enabling to clean the data around GW200129 using the linear subtraction tool \texttt{gwsubtract}~\citep{o3b_cat,vetoes_report, rf_alog}.

To estimate the effectiveness of linear subtraction, we compared the data from the original (i.e.\,not cleaned) LIGO Livingston data frame and the frame that has the excess noise removed using \texttt{gwsubtract}. 

First, we identified when RF noise was present in the data around GW200129 using Z-score.
Out of $4096$\,s, we selected in total $366$\,s considered to contain radio frequency noise.
After that, the data subset was Q-tiled for various $Q$ and $f_0$ values.

\begin{figure}
    \centering
    \begin{subfigure}[b]{0.4\paperwidth}
        \centering
        \includegraphics[width=0.4\paperwidth]{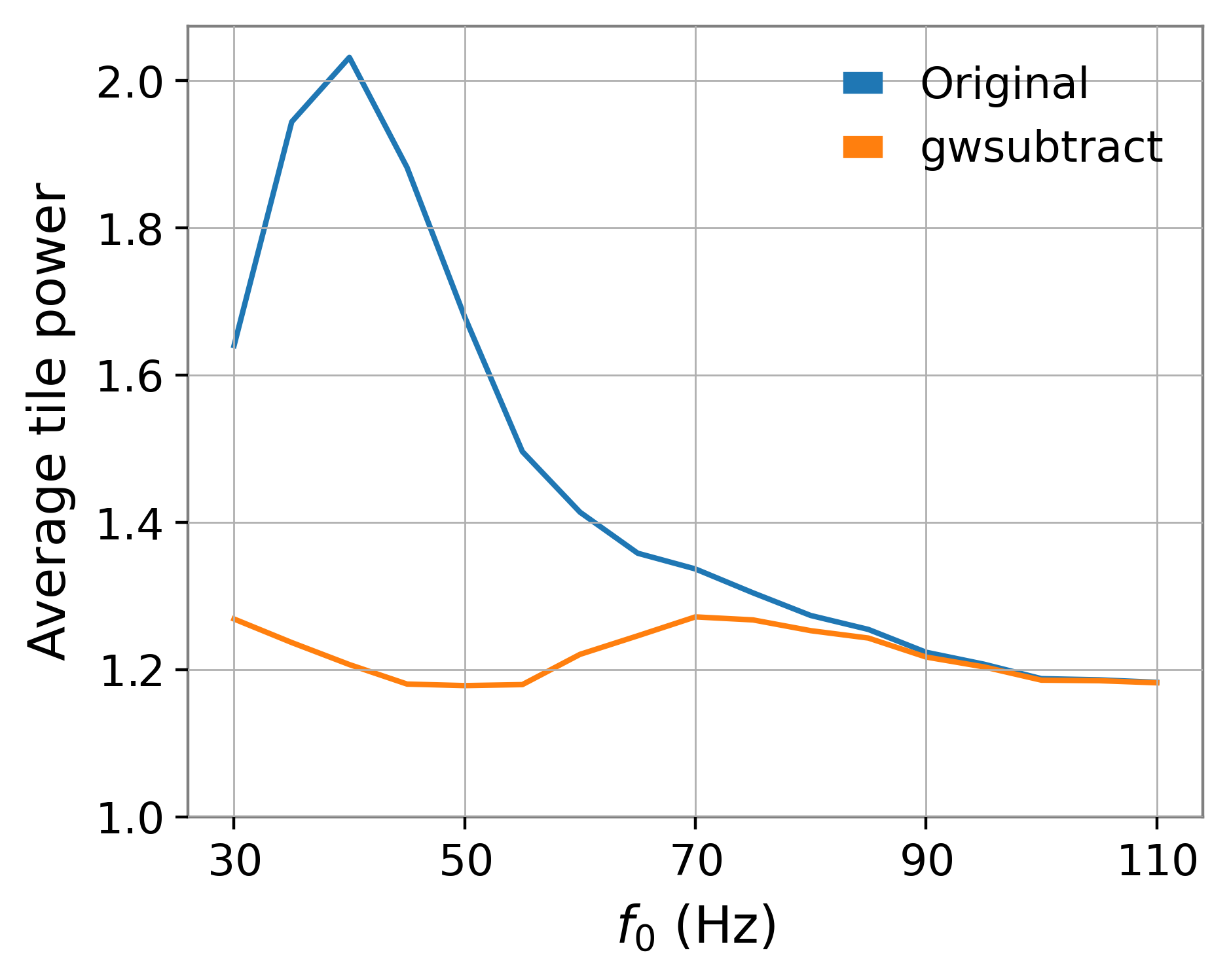}
        \caption{Average tile power for data around GW200129 with quality factor $Q=8$.}\label{fig:gw200129_avg_tile_power}
    \end{subfigure}
    \begin{subfigure}[b]{0.4\paperwidth}
        \centering
        \includegraphics[width=0.4\paperwidth]{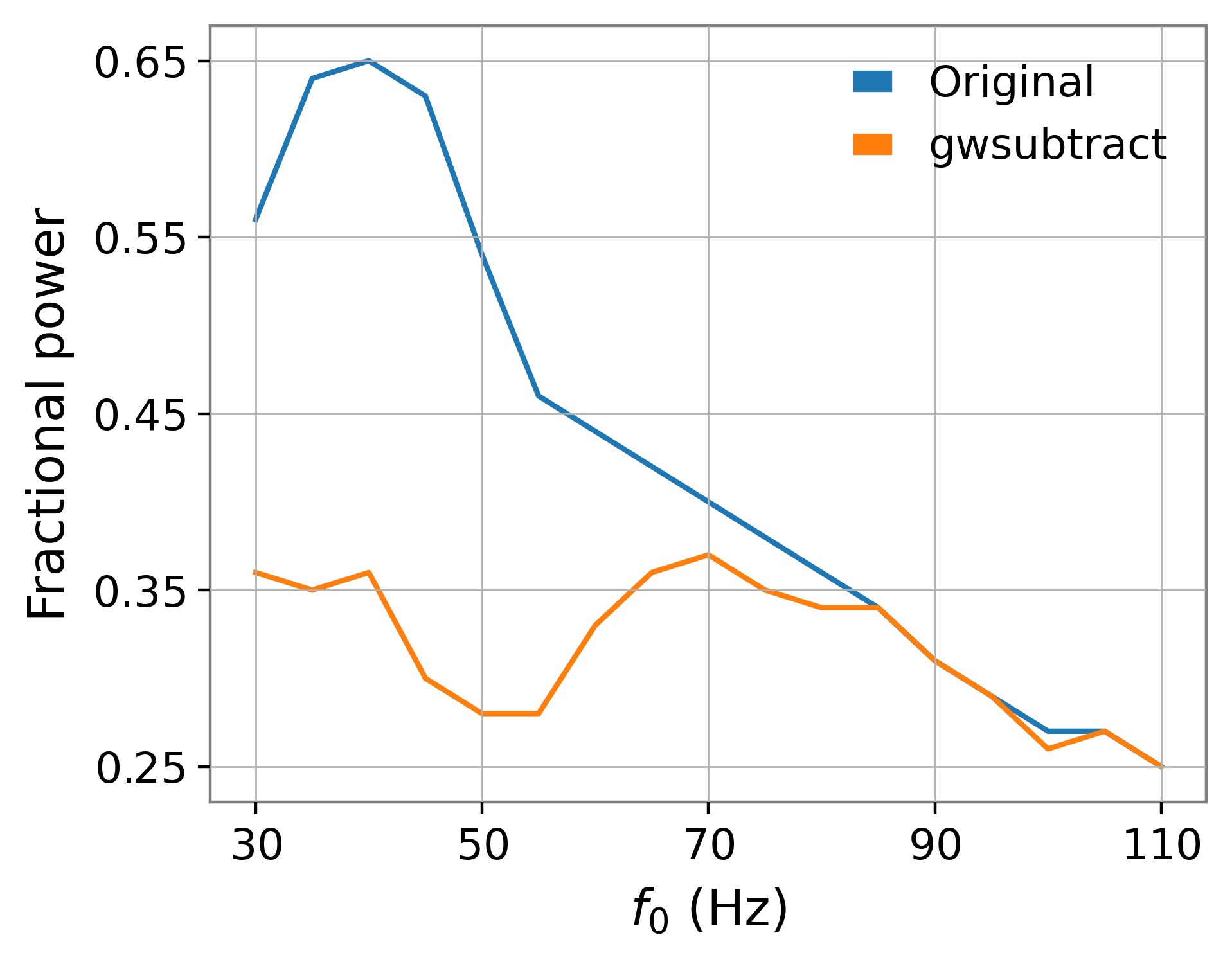}
        \caption{Fractional power for data around GW200129 with quality factor $Q=8$.}\label{fig:gw200129_fpower}
    \end{subfigure}
    \caption{Average tile power (above) and fractional power (below) for data around GW200129 with quality factor $Q=8$. 
    Linear subtraction \texttt{gwsubtract} removes a significant portion of the non-Gaussian noise below 50\,Hz, yet some of the noise still remains after the cleaning. At frequencies above $\sim 85$\,Hz, there is no excess noise removed.}\label{fig:gw200129}
\end{figure}

For the first test of non-Gaussianity, we simply estimate what is the average tile power for both the original and the \texttt{gwsubstract} data.
To do that, we Q-tile the data for various Q and $f_0$ values which give us the total number of tiles and the power in each tile. 
Since we use the normalised Q-tiles, the average tile power for Gaussian data is 1.
Figure \ref{fig:gw200129_avg_tile_power} shows the corresponding average tile power for quality factor $Q=8$.
We chose the Q-value of 8 in this and the following figures because low Q-value tiles fit the short duration glitches such as RF noise better.

As a more advanced test, we want to measure the amount of non-Gaussianity.
In order to do that, we model the data with an exponential distribution with a rate $\lambda=1/2$  and a half Student's T distribution~\cite{student_t} with probability density function given by:
\begin{equation}
    \label{eq:half_student}
    f(x | \sigma, \nu) = \frac{2\Gamma(\frac{\nu+1}{2})}{\Gamma(\frac{\nu}{2})\sqrt{\nu \pi \sigma^2}} \bigg(1 + \frac{1}{\nu}\frac{x^2}{\sigma^2} \bigg)^{-\frac{\nu+1}{2}}
\end{equation}
where $\nu$ is degrees of freedom, $\sigma$ is normality parameter and $\Gamma$ is the gamma function.
We chose half Student's T distribution because it has a long tail which closely matches the non-Gaussian data (see, for example, Fig.~\ref{fig:power_hists}).

After recovering distribution parameters, we estimate the fractional power which is the ratio of non-Gaussian power to the total power. Figure \ref{fig:gw200129_fpower} shows the fractional power for GW200129.
\subsection{Light scattering noise}
\label{sub:sc}
We also apply both tests to measure the excess noise in the time-series data that is continuous.
For this example, we selected 10 minutes of data with a relatively low amount of light scattering noise and 10 minutes of scattering-heavy data recorded by LIGO Livingston on January 6th, 2020.  
Light scattering glitches are caused by stray light reflection in the beamtube, which can happen due to excessive ground motion\,~\citep{Soni_2021, accadia_2010, Tolley_2023}.
As a result, such noise transients can last up to minutes with a typical frequency range of $20\text{--}50$\,Hz.

\begin{figure}
    \centering
    \begin{subfigure}[b]{0.4\paperwidth}
        \centering
        \includegraphics[width=0.4\paperwidth]{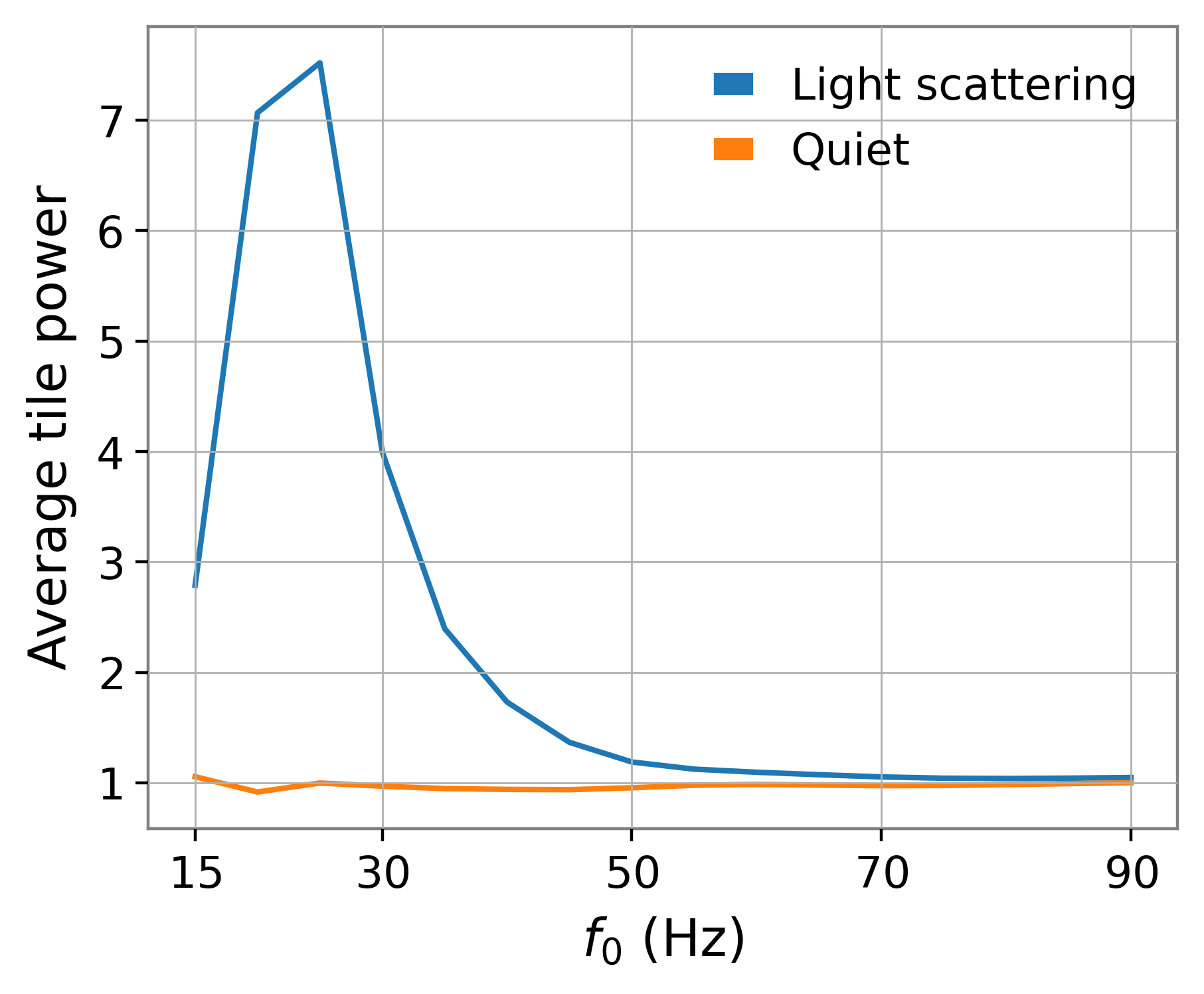}
        \caption{Average tile power for data with quality factor $Q=8$ in the presence and absence of light scattering artefacts.}\label{fig:scs_avg_tile_power}
    \end{subfigure}
    \begin{subfigure}[b]{0.4\paperwidth}
        \centering
        \includegraphics[width=0.4\paperwidth]{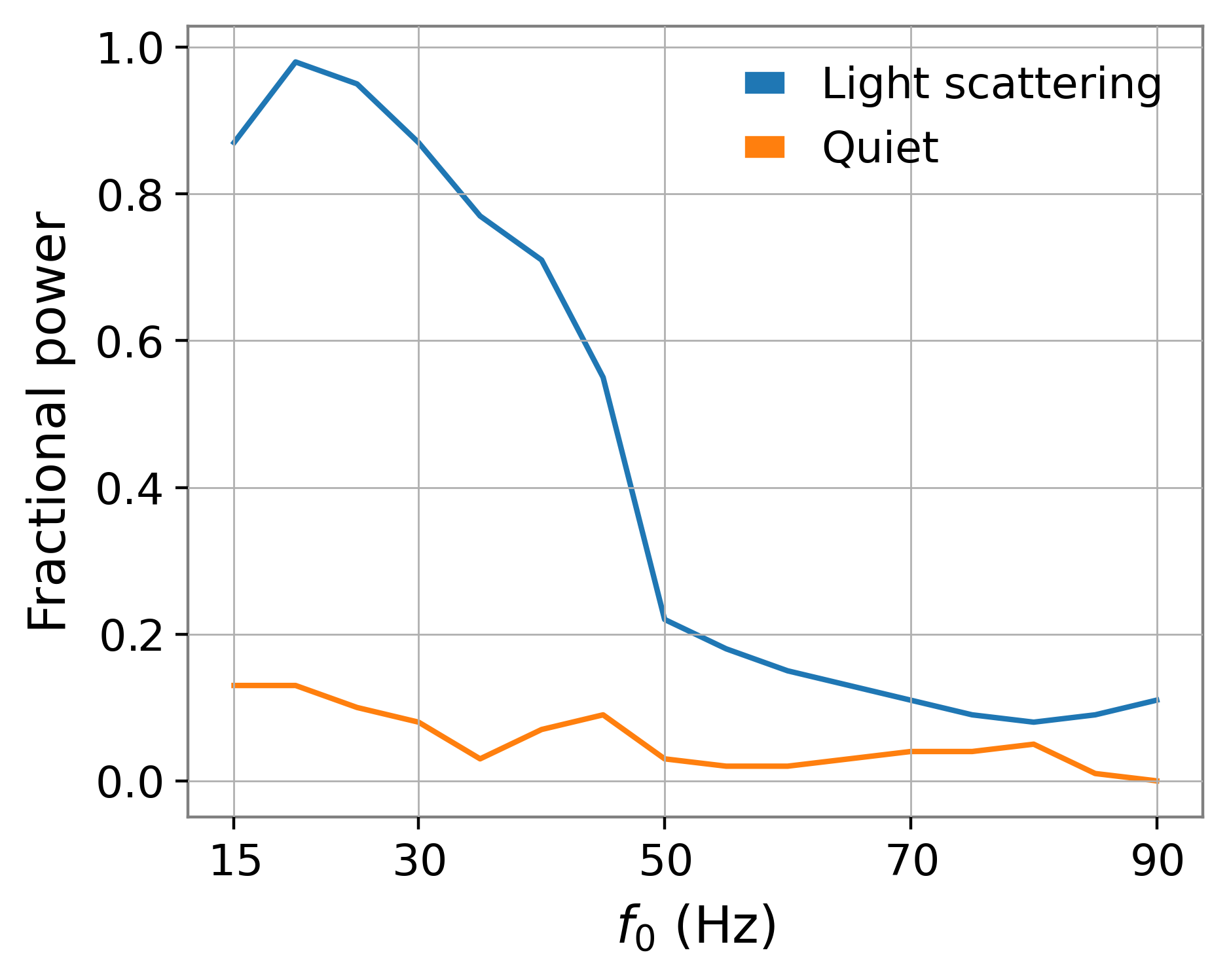}
        \caption{Fractional power for data with quality factor $Q=8$ in the presence and absence of light scattering artefacts.}\label{fig:scs_fpower}
    \end{subfigure}
    \caption{Average tile power (above) and fractional power (below) for data with quality factor $Q=8$.
    Time series data in blue has excess power at the 15--50\,Hz range caused by the light scattering artefacts, whereas the time series data in orange with no light scattering artefacts have a relatively small amount of non-Gaussian noise across all frequencies.
    }\label{fig:scs}
\end{figure}

We use the same procedure as in the previous example of GW200129.
Firstly, we calculate the Q-tile power for specific $Q$ and $f_0$ values giving us the average tile power.
Then, we fit the Q-tile power with $\chi^2(2)$ and half Student's T distributions to estimate the fractional power which is the ratio of non-Gaussian power to the total power. Figure \ref{fig:scs} shows the average tile power and the fractional power for both low noise and scattering-heavy time series using quality factor $Q=8$.
\section{Discussion}
\label{sec:disc}

In this paper, we present a method to estimate the amount of non-Gaussian noise in the data based on the normalised Q-transform.
This method has multiple advantages over other statistical tests.
While comparing power spectral density (PSD) between the cleaned and original data provides a quantitative measurement of the removed power, it does not discriminate whether the removed power is non-Gaussian, i.e.~the excess noise, or Gaussian~\cite{o2_detchar}.
Furthermore, the PSD estimation is usually performed over a long duration, e.g.~512\,s, meaning that such a test becomes less sensitive to removing short-duration noise bursts like in the data around GW200129.

Contrary to the PSD comparison, our test discriminates between the Gaussian and non-Gaussian noise; we also analyze only the times when the RF noise is present which makes our test much more sensitive than the PSD comparison.

Another commonly used method to estimate how well the data is cleaned relies on the visual inspection of spectrograms~\cite{o2_detchar}.
However, this becomes problematic when the excess noise is weak and is comparable to Gaussian noise.
This issue becomes even more precarious when the excess noise overlaps a GW signal, as in the case of GW200129. 
Conversely, our method provides a quantitative measurement of the excess noise.

By using normalised Q-tiles, we put a constraint that Gaussian data has an average tile power of 1, whereas the data with excess noise is expected to have more power on average.
Furthermore, we found that the Gaussian data must follow a $\chi^2(2)$ distribution after we normalise the Q-tiles.
As a result, the data containing Gaussian and non-Gaussian noise can be decomposed into two distributions, one of which is $\chi^2(2)$.

To estimate the amount of non-Gaussian noise in GW data, we fit the normalised Q-transform power using Bayesian statistical modelling with Hamiltonian Monte Carlo.
By doing this, we reconstruct the parameters of both Gaussian and non-Gaussian distributions.

As the first application of this method, we investigate if (and how much) \texttt{gwsubtract} cleaned the data around GW200129.
As shown in  Figure \ref{fig:gw200129}, RF noise is mostly affecting the data in $30\text{--}50$\,Hz range where the non-Gaussian power is at least 55\% of the total data power, i.e.\,the frequencies where precession effects are especially important for GW200129~\cite{Payne_2022}.
Linear subtraction \texttt{gwsubtract} removes up to 30\% of the fractional power in this frequency range indicating that it does remove the RF noise, albeit not entirely.

At higher frequencies, there is less RF noise in the data which could explain why \texttt{gwsubtract} is less efficient at these frequencies compared to the $30\text{--}50$\,Hz range.
Above $\sim 85$\,Hz, \texttt{gwsubtract} does not remove any noise which is reasonable given that the witness channel used to clean the data does not have any information above this frequency.

Our simpler test showing the average tile power agrees well with the fractional power results (Figure \ref{fig:gw200129}).
Both frames, Original and \texttt{gwsubtract}, have more power per tile than the Gaussian data. However, the original data has considerably more power in the $30\text{--}50$\,Hz range (up to 70\% more). 

We also tested if our method can be applied as a live monitoring tool of excess noise in GW detector data. 
For that, we compared 10 minutes of relatively quiet data with 10 minutes of data containing many scattering glitches at LIGO Livingston on January 6th, 2020. 

We found that the fractional power is much higher for data with many scattering glitches than for the quiet data (Figure \ref{fig:scs_fpower}).
The excess noise is concentrated within $10\text{--}50$\,Hz and peaks at $20$\,Hz, where the non-Gaussian noise power is 0.96 of the total power.
The frequency of light scattering noise ($20\text{--}50$\,Hz) is similar to the excess noise at $15\text{--}50$\,Hz, indicating that our test correctly identified the presence of excess noise.
For 10 minutes of relatively low noise, the fractional power is much lower and stays relatively constant across all frequencies.

Similarly, our average tile power test indicates that the data with scattering glitches contains up to $\sim7$ times more power than the Gaussian data (Figure \ref{fig:scs_avg_tile_power}, blue).
Data with no scattering glitches (Figure \ref{fig:scs_avg_tile_power}, orange) stays relatively constant across all frequencies and has an average tile power close to Gaussian data, i.e.\,1.

\section{Conclusions}
\label{sec:conc}

Gravitational-wave data contains many noise transients which can affect GW searches and estimation of source parameters such as the precession of a binary black hole. 
It is possible to mitigate excess noise from GW data in some cases, for example by correlating information from noise witness channels. 
However, the lack of meaningful and sensitive statistical tests prevents estimating the effectiveness of such noise mitigation tools.

In this paper, we propose a novel method to statistically measure the non-Gaussian noise in GW detector data.
By using normalised Q-transform, we constrain how Gaussian data should be distributed, which allows us to determine the amount of non-Gaussian noise in the data.

Our method provides two statistical tests to measure the non-Gaussianity.
The first one measures the average tile power which, for Gaussian data, must be equal to 1.
The second test estimates the relative power of non-Gaussian data compared to the total power, which we define as \textit{fractional power}.
Both our tests are sensitive to excess noise changes in frequency, which allows for a more precise noise measurement than averaging across all frequencies.

In order to show the effectiveness of our statistical tests, we explore two scenarios.
First, we investigate how well \texttt{gwsubtract}, a linear subtraction tool, removes the noise around the binary black hole event GW200129.
We find that \texttt{gwsubtract} removes a significant portion of the non-Gaussian noise at lower frequencies (30--70\,Hz), yet it fails to remove the non-Gaussian noise completely. There is also no noise removed above $\sim 85$\,Hz.

In the second scenario, we test if our method can be applied as a live monitoring tool of excess noise in GW detector data.
We show that for data that has many light scattering noise artefacts, our method correctly identifies the excess noise within the light scattering noise frequency range (20--50\,Hz).

\begin{acknowledgments}

We thank Marissa Walker and members of the LIGO Detector Characterization group for their valuable input during the preparation of this manuscript.
We also thank the anonymous referee for valuable comments which improved the manuscript during the review.

RM is supported by STFC grants ST/S000550/1, ST/T000325/1, ST/V005715/1 and ST/X002225/1.
The authors are grateful for computational resources provided by the LIGO Laboratory and supported by National Science Foundation Grants PHY-0757058 and PHY-0823459.

This research has made use of data, software and/or web tools obtained from the Gravitational Wave Open Science Center (https://www.gw-openscience.org), a service of LIGO Laboratory, the LIGO Scientific Collaboration, the Virgo Collaboration, and KAGRA.
This material is based upon work supported by NSF’s LIGO Laboratory which is a major facility fully funded by the National Science Foundation. 
LIGO Laboratory and Advanced LIGO are funded by the United States National Science Foundation (NSF) as well as the Science and Technology Facilities Council (STFC) of the United Kingdom, the Max-Planck-Society (MPS), and the State of Niedersachsen/Germany for support of the construction of Advanced LIGO and construction and operation of the GEO600 detector.
Additional support for Advanced LIGO was provided by the Australian Research Council.
Virgo is funded, through the European Gravitational Observatory (EGO), by the French Centre National de Recherche Scientifique (CNRS), the Italian Istituto Nazionale di Fisica Nucleare (INFN) and the Dutch Nikhef, with
contributions by institutions from Belgium, Germany, Greece, Hungary, Ireland, Japan, Monaco, Poland, Portugal, Spain.
KAGRA is supported by Ministry of Education, Culture, Sports, Science and Technology (MEXT), Japan Society for the Promotion of Science (JSPS) in Japan; National Research Foundation (NRF) and Ministry of Science and ICT (MSIT) in Korea; Academia Sinica (AS) and National Science and Technology Council (NSTC) in Taiwan.

Various parts of the analysis used \texttt{GWpy} \cite{gwpy}, \texttt{PyCBC} \citep{pycbc, pycbc_live_o3} \texttt{PyMC} \cite{pymc}, 
\texttt{NumPy} \cite{numpy}, \texttt{SciPy} \cite{scipy}, \texttt{IPython} \cite{ipython}, \texttt{Jupyter notebook} \cite{jupyter}, \texttt{pandas} \cite{pandas}, \texttt{Matplotlib} \cite{matplotlib}, \texttt{arviz} \cite{arviz} and \texttt{corner} \cite{corner}. 

This document has been assigned LIGO Laboratory document number P2300172.

For the purpose of open access, the author has applied a Creative Commons Attribution (CC BY) licence to any Author Accepted Manuscript version arising.
\end{acknowledgments}

\clearpage
\bibliography{refs_papers,refs_software}
\end{document}